\documentclass[11pt,a4paper]{article}
\usepackage{jheppub}
\usepackage{amsmath}
\usepackage[most]{tcolorbox}
\usepackage{dsfont}
\usepackage{ulem}
\usepackage{natbib}
\usepackage{xcolor}
\usepackage[hang,flushmargin]{footmisc}
\usepackage{tikz-cd}
\usepackage{enumitem}
\usepackage{amsfonts,amssymb, amscd,amsmath,latexsym,amsbsy,bm}
\usepackage{stmaryrd}
\usepackage{todonotes}
\usepackage{float}

\usepackage{romannum}

\makeatletter\renewcommand{\@biblabel}[1]{#1.}\makeatother

\newtcolorbox{empheqboxed}{colback=gray!20, 
 colframe=white,
 width=\textwidth,
 sharpish corners,
 top=0mm, 
 bottom=0pt
}

\title{Flipping relation as a reduced star-star relation}
\author{Erdal Catak$^a$ and Mustafa Mullahasanoglu$^{b,c}$}
\affiliation{
$^{a}$ Department of Physics, Istanbul University,\\ 34134 Istanbul, Türkiye \\[-0.4cm]

$^b$ Department of Physics, Bogazici University,\\ 34342 Bebek, Istanbul, Türkiye\\[-0.4cm]

$^c$ Feza Gursey Center for Physics and Mathematics, Bogazici University,\\ 34684, Kandilli,
Istanbul, Türkiye
}

\emailAdd{ecatak@istanbul.edu.tr}
\emailAdd{mustafa.mullahasanoglu@std.bogazici.edu.tr}

\abstract{
In this paper, we consider the lens hyperbolic gamma solution to the star-star relation and the flipping relation from three-dimensional $\mathcal{N}=2$ supersymmetric gauge theories on $S^3_b/\mathbb{Z}_r$. We explore that a certain limit of the star-star relation yields the latter symmetry transformation, which exchanges the edge interactions of two outer spins with a centrally sited spin. Furthermore, we obtain more solutions to the flipping relation in terms of the hyperbolic gamma, basic hypergeometric, and the Euler gamma functions.
}

\keywords{Star-star relation, gauge/YBE correspondence, dual supersymmetric gauge theories, flipping relation, lattice spin models, integrability.}

\begin{document}

\maketitle

\section{Introduction}

Over the past decade, the gauge/YBE correspondence has involved numerous solutions to the integrability condition, the star-star relation \cite{Baxter:1997ssr}, for lattice spin models in statistical mechanics. The most general solution, constructed by Masahito Yamazaki \cite{Yamazaki:2013nra}, covers most of the known integrable models in appropriate limits. Further solutions in terms of hypergeometric gamma functions have been derived via the gauge/YBE correspondence  \cite{Catak:2021coz}. In this framework \cite{Gahramanov:2017ysd, Yamazaki:2018xbx, Yagi:2016oum}, the partition functions of dual supersymmetric gauge theories produce integral identities, which are interpreted as integrability conditions.

In this work, we revisit the lens hyperbolic solution to the star-star relation \cite{Mullahasanoglu:2021xyf,Mullahasanoglu:2024gib} and the flipping relation \cite{Catak:2024ygo}. We demonstrate that the latter relation is a reduced form of the integrability condition: the star-star relation. 
Both relations are the result of the partition functions of three-dimensional $\mathcal{N}=2$ supersymmetric gauge theories on $S^3_b/\mathbb{Z}_r$ \cite{Benini:2011nc, Imamura:2012rq, Imamura:2013qxa}. 

By reducing two diagonal spins in a specific limit of the star-star relation, we obtain a further symmetry of the model. This resulting relation allows for the exchange of edge interactions between two outer spins and a central spin. From the perspective of supersymmetric gauge theory, this corresponds to a reduction in the flavor symmetry of the dual theories. We achieve this symmetry reduction at the level of partition functions by taking limits of the fugacities, a procedure justified by the asymptotic properties of the lens hyperbolic gamma function.

We also investigate that this property extends to other integrable models obtained by the gauge/YBE correspondence. So we solve the flipping relation in terms of the hyperbolic gamma \cite{Spiridonov:2010em}, the basic hypergeometric function  \cite{Gahramanov:2016wxi, Gahramanov:2013rda, Gahramanov:2014ona, Gahramanov:2015cva, Catak:2022glx}, and the complex Euler gamma function \cite{Kels:2013ola, Kels:2015bda, Eren:2019ibl}. 

The organization of the paper is as follows.
In Section 2, we deduce a flipping relation by reducing the star-star relation. We take the limit from the partition functions of the supersymmetric theories on the lens squashed three-sphere. In Section 3, we solve the flipping relation in terms of the hyperbolic gamma, the basic hypergeometric function, and the complex Euler gamma functions.

\section{The flipping relation as a reduced star-star relation}
Recent works have shown that integrable lattice spin models derived from the gauge/YBE correspondence can be decorated or mapped to various dual lattices using a symmetry transformation of the Ising model \cite{Mullahasanoglu:2023nes, Catak:2024ygo}. One such transformation is the decoration transformation \cite{Fisher1959, Syozi:1980iw}, a symmetry that introduces a new spin at the center of a nearest-neighbor interaction. While seemingly trivial, applying this transformation to a hexagonal lattice and subsequently eliminating the original spins via the star-triangle relation yields a non-trivial Kagome lattice.

In this work, we demonstrate that these integrable models also possess a distinct symmetry transformation known as the flipping relation. We examine this relation from three perspectives: its role in generating decorated models, its significance for one-dimensional models, and its connection to the integrability condition.

First, the flipping relation provides a criterion for the decoration transformation. This transformation introduces a new spin into a nearest-neighbor interaction, generating two new edge interactions between the original spins and the central one. The key consideration is the nature of these generated interactions. If the Boltzmann weights (governing the interactions between the original spins and the new central spin) satisfy the flipping relation, it guarantees that the outcome of the decoration is identical in both configurations. In other words, the two new edge interactions are created in a way that makes them functionally equivalent. This property establishes the freedom to apply the decoration transformation. The integrable models we consider will have this freedom in the sense of exchanging spectral parameters.

Second, the flipping relation has a direct interpretation in one-dimensional models. The partition function of a one-dimensional model can be evaluated via the transfer matrix method if its Boltzmann weights commute. The flipping relation is precisely this commutation condition. Therefore, we conclude that any one-dimensional model obeying the flipping relation possesses commuting transfer matrices, enabling the computation of its partition function.

We further demonstrate that the flipping relation can be derived as a specific reduction of the star-star relation. This connection is significant because certain models are known to satisfy the star-star relation without satisfying the star-triangle relation, such as the three-layer Zamolodchikov model \cite{Baxter:1997ssr}. An analogous difference exists between the flipping relation and the decoration transformation. Specifically, the Boltzmann weights of the three-layer Zamolodchikov model satisfy the flipping relation\footnote{This can be verified by fixing the spectral parameters $\beta=\delta=0$ in equation (6) of \cite{Baxter:1997ssr}, using definition (11).} but do not satisfy the decoration transformation.
However, in Interaction-Round-a-Face (IRF) and vertex models\footnote{For the construction of vertex models from edge interaction models via the Bailey lemma, see \cite{Gahramanov:2015cva, Gahramanov:2022jxz}.}, the flipping relation expresses the self-commutativity of transfer matrices \cite{BAXTER1986321}.
 
The mathematical expression of the flipping relation for a model with spins $\sigma_i=(x_i,m_i)$ taking values from $ \mathbb{R}\times \mathbb{Z}$ or its subsets ($x_i\in \mathbb{R}$ continuous spin variables and  $m_i\in \mathbb{Z}$ discrete spin variables) is the following equality
\begin{align}
   \sum_{m_0} \int dx_0\: S(\sigma_0) W_{\alpha_1,\beta_1}(\sigma_1,\sigma_0)&W_{\alpha_2,\beta_2}(\sigma_2,\sigma_0)
   \nonumber \\ &
   = \sum_{y_0} \int dz_0\: S(\sigma_0) W_{\alpha_2,\beta_2}(\sigma_1,\sigma_0)W_{\alpha_1,\beta_1}(\sigma_2,\sigma_0)
   \label{flippingdefinition},
\end{align}
where  $W_{\alpha,\beta}(\sigma_i,\sigma_j)$ is the Boltzmann weight of nearest neighbor interaction with $\alpha,\: \beta$ continuous and discrete spectral parameters, respectively. $S(\sigma_i)$ is a self-interaction, independent of the spectral parameters, for each spin variable and can be read as an external field contribution.

\subsection{Revisiting lens hyperbolic solution}

Recently, in \cite{Catak:2024ygo}, the flipping relation was solved in terms of the lens hyperbolic gamma functions\footnote{The lens hyperbolic gamma function \cite{Gahramanov:2016ilb} is defined as a combination of hyperbolic gamma functions
\begin{align}
\gamma_h(z,y;\omega_1,\omega_2) 
=    \gamma^{(2)}(-iz-i\omega_1y;-i\omega_1r,-i\omega) \times \gamma^{(2)}(-iz-i\omega_2(r-y);-i\omega_2r,-i\omega) \:,
\end{align}
where $\text{Im}(\omega_1/\omega_2)>0$, $r\in\{1,2,...\}$, $y\in \{0,1,...,r-1\}$, and  $\omega:=\omega_1+\omega_2$.

We introduce the $q$-Pochhammer symbol $(z;q)_{\infty}=\prod_{i=0}^{\infty}(1-zq^i)$ and use the shorthand notation $(z,x;q)_\infty=(z;q)_\infty(x;q)_\infty$. Then we define the hyperbolic gamma function 
\begin{align}
	\gamma^{(2)}(z;\omega_{1},\omega_{2})=e^{\frac{\pi i}{2}B_{2,2}(z;\omega_{1},\omega_{2})}\frac{(e^{-2\pi i\frac{z}{\omega_{2}}}\tilde{q};\tilde{q})_\infty}{(e^{-2\pi i\frac{z}{\omega_{1}}};q)_\infty} \; ,
\end{align}
where $\tilde{q}=e^{2\pi i \omega_{1}/\omega_{2}}$, $q=e^{-2\pi i \omega_{2}/\omega_{1}}$
and the Bernoulli polynomial is 
$
 B_{2,2}(z;\omega_1,\omega_2)=\frac{z^2-z\omega}{\omega_1\omega_2}+\frac{\omega^2+\omega_1\omega_2}{6\omega_1\omega_2}$.
}. The corresponding integral identity represents the duality of supersymmetric gauge theories possessing $SU(2)$ gauge symmetry and $SU(4)$ flavor symmetry.  The equality of partition functions is the following integral transformation formula
\begin{align}
      \frac{1}{2r\sqrt{-\omega_1\omega_2}} &\sum_{y=0}^{[ r/2 ]} \int _{-\infty}^{\infty} dx\frac{\prod_{i=1}^4\gamma_h(a_i\pm z,u_i\pm y;\omega_1,\omega_2)}{\gamma_h(\pm 2z,\pm 2y;\omega_1,\omega_2)}
\nonumber\\ 
&=\frac{\gamma_h(a_1+a_2,u_1+u_2;\omega_1,\omega_2)\gamma_h(a_3+a_4,u_3+u_4;\omega_1,\omega_2)}{\gamma_h(\Tilde{a}_1+\Tilde{a}_2,\Tilde{u}_1+\Tilde{u}_2;\omega_1,\omega_2)\gamma_h(\Tilde{a}_3+\Tilde{a}_4,\Tilde{u}_3+\Tilde{u}_4;\omega_1,\omega_2)}
\nonumber\\ &\times
      \frac{1}{2r\sqrt{-\omega_1\omega_2}} \sum_{m=0}^{[ r/2 ]} \int _{-\infty}^{\infty} dx\frac{\prod_{i=1}^4\gamma_h(\Tilde{a}_i\pm x,\Tilde{u}_i\pm m;\omega_1,\omega_2)}{\gamma_h(\pm 2x,\pm 2m;\omega_1,\omega_2)}
      \label{flippingequation1}
\end{align}
where fugacities on the right-hand side are shifted by $s$ and $p$ chemical potentials
\begin{align}
    \Tilde{a}_{1,2}=a_{1,2}+s \:,\quad
\Tilde{a}_{3,4}=a_{3,4}-s \:,\quad
  \Tilde{u}_{1,2}= u_{1,2}+p \:,\quad
\Tilde{u}_{3,4}=u_{3,4}-p \:.
\label{tilde1}
\end{align}
Then one can see that the integral identity (\ref{flippingequation1}) is a symmetry property of the following Boltzmann weight of the  integrable lens hyperbolic model \cite{Gahramanov:2016ilb}
\begin{align}
\begin{aligned} \label{boltamannweightSU2}
    W_{\alpha_i,\beta_i}(x_i,x_j,m_i,m_j) &=    
    \gamma_h(-\alpha_i\pm x_i\pm x_j,-\beta_i\pm m_i\pm  m_j;\omega_{1},\omega_2)    \;,
\end{aligned}
	\end{align}
    under the change of variables
\begin{align}
    a_{1,2}=-\alpha_1\pm x_1 \:,\quad
a_{3,4}=-\alpha_2\pm x_2\:,\quad
 u_{1,2}= -\beta_1\pm m_1 \:,\quad
u_{3,4}= -\beta_2\pm m_2\:,
\label{tilde1}
\end{align}
and 
\begin{align}
    s=\alpha_1-\alpha_2 \:,\quad
p=\beta_1-\beta_2\:.
\label{sp}
\end{align}
Then the flipping relation (\ref{flippingdefinition}) is satisfied by the Boltzmann weight (\ref{boltamannweightSU2}). 

In this work, our goal is to show that the flipping relation is a reduction of the star-star relation 
\begin{equation}
R\left(\begin{array}{cc}
    \sigma_4     & \sigma_3 \\
\sigma_1         & \sigma_2
    \end{array}\right)
=\frac{W_{2\eta-\alpha_3-\alpha_4}(\sigma_1,\sigma_2)W_{2\eta-\alpha_2-\alpha_3}(\sigma_1,\sigma_4)}{W_{2\eta-\alpha_2-\alpha_3}(\sigma_2,\sigma_3)W_{2\eta-\alpha_3-\alpha_4}(\sigma_3,\sigma_4)}
\overline{R}\left(\begin{array}{cc}
    \sigma_4    & \sigma_3 \\
\sigma_1         & \sigma_2
    \end{array}\right),
    \label{starstar}
\end{equation}
where we have a condition on the spectral parameters $\sum_{i=1}\alpha_i=\eta$ and $\sum_{i=1}\beta_i=\theta$, and we omit discrete spectral parameters appearing in the same form just for the simplification. The star Boltzmann weights are defined to remind the IRF models 
\begin{equation}
R\left(\begin{array}{cc}
\sigma_{4} & \sigma_{3} \\
\sigma_{1} & \sigma_{2}
\end{array}\right)=\sum_{m_0} \int dx_0  W_{\alpha_1,\beta_1}\left(\sigma_{1}, \sigma_{0}\right) W_{\alpha_2,\beta_2}\left(\sigma_{0}, \sigma_{2}\right) W_{\alpha_3,\beta_3}\left(\sigma_{3}, \sigma_{0}\right) W_{\alpha_4,\beta_4}\left(\sigma_{0}, \sigma_{4}\right),
\label{factorized1}
\end{equation}
and
\begin{equation}
\overline{R}\left(\begin{array}{cc}
\sigma_{4} & \sigma_{3} \\
\sigma_{1} & \sigma_{2}
\end{array}\right)=\sum_{m_0} \int dx_0  W_{\alpha_3,\beta_3}\left(\sigma_{1}, \sigma_{0}\right) W_{\alpha_4,\beta_4}\left(\sigma_{0}, \sigma_{2}\right) W_{\alpha_1,\beta_1}\left(\sigma_{3}, \sigma_{0}\right) W_{\alpha_2,\beta_2}\left(\sigma_{0}, \sigma_{4}\right).
\label{factorized2}
\end{equation}

Now we consider the integral identity obtained in \cite{Mullahasanoglu:2021xyf, Mullahasanoglu:2024gib} in the same class of supersymmetric dualities. That is, the star-star relation (\ref{starstar}) solved by the Boltzmann weights, (\ref{boltamannweightSU2}) and the following integral identities are equivalent\footnote{More specifically, the star-star relation (\ref{starstar}) can be obtained by permuting and applying the transformation formula (\ref{SSR-s3bZr}) twice for a certain choice of parameters. Then the star-star relation will be obtained when the variables are changed in the result
\begin{align}
    \begin{array}{c}
 a_{1,2}=\pm x_1-\alpha_1\,, \quad a_{3,4}=\pm x_3-\alpha_3\,, \quad  u_{1,2}=\pm m_1-\beta_1\,, \quad u_{3,4}=\pm m_3-\beta_3\,, \\
   a_{5,6}=\pm x_2-\alpha_2\,, \quad a_{7,8}=\pm x_4-\alpha_4\,, \quad  u_{5,6}=\pm m_2-\beta_2\,, \quad u_{7,8}=\pm m_4-\beta_4\:.
\end{array}
\label{chngvs}
\end{align}
with the balancing conditions 
\begin{align}  \sum_{i=1}^8a_i\equiv -2\sum_{i=1}^4\alpha_i= -4\eta\equiv 2(\omega_1+\omega_2)\:, \quad 
\sum_{i=1}^8u_i\equiv -2\sum_{i=1}^4\beta_i= -4\theta\equiv 0\:.
\end{align}} in some sense \cite{Dolan:2008qi, Gahramanov:2013xsa}
\begin{equation} \begin{aligned}
          \sum_{m=0}^{[ r/2 ]} \epsilon (m)\int _{-\infty}^{\infty} &\frac{\prod_{i=1}^8\gamma_h(a_i\pm x,u_i\pm m;\omega_1,\omega_2)}{\gamma_h(\pm 2x,\pm 2m;\omega_1,\omega_2)} 
     \frac{dx}{r\sqrt{-\omega_1\omega_2}}
      \\ 
     =&
     \frac{\prod_{1\leq i<j\leq 4}\gamma_h(a_i + a_j,u_i + u_j;\omega_1,\omega_2) }
     {\prod_{5\leq i<j\leq 8}\gamma_h(\Tilde{a}_i+\Tilde{a}_j,\Tilde{u}_i+\Tilde{u}_j;\omega_1,\omega_2)}  \\
     & \times   \sum_{y=0}^{[ r/2 ]}\epsilon (y) \int _{-\infty}^{\infty} 
\frac{\prod_{i=1}^8\gamma_h(\Tilde{a}_i\pm z,\Tilde{u}_i\pm y;\omega_1,\omega_2)}{\gamma_h(\pm 2z,\pm 2y;\omega_1,\omega_2)} 
     \frac{dz}{r\sqrt{-\omega_1\omega_2}}
      \:,\label{SSR-s3bZr}
      \end{aligned} 
\end{equation}
where the balancing conditions are $\sum_{i=1}^8a_i=2(\omega_1+\omega_2)$, $\sum_{i=1}^8u_i=0$,
and the chemical potentials are defined as
\begin{equation}
\begin{aligned}
s & =\frac{1}{2}(\omega_1+\omega_2-\sum_{i=1}^4a_i)=\frac{1}{2}(-\omega_1-\omega_2+\sum_{i=5}^8a_i) \\
p & =-\frac{1}{2}\sum_{i=1}^4u_i=\frac{1}{2}\sum_{i=5}^8u_i\:,
\end{aligned}
\end{equation}
for shifting the fugacities 
\begin{equation}
\begin{aligned}
     \tilde{a}_i & =  a_i+s, & \tilde{u}_i & =u_i+p,  & \text{if} \;\;\; i=1,2,3,4 \:,
     \\ 
     \tilde{a}_i & =  a_i-s,  & \tilde{u}_i & = u_i-p, & \text{if} \;\;\; i=5,6,7,8.
\end{aligned}
\end{equation}

Finally, we show that the integral identity, star-star relation, (\ref{SSR-s3bZr}) can be reduced to the integral identity, the flipping relation, (\ref{flippingequation1}) by the following limits
\begin{align}
\begin{aligned}
    a_3 & \to \infty  \: , & a_4 & =\omega_1+\omega_2-2 s-\sum_{i=1}^3a_i \to -\infty  ~~~ \text{and} & u_3 & = u_4\: ,
    \\
     a_7 & \to \infty  \: , & a_8 & =\omega_1+\omega_2+2 s-\sum_{i=5}^7a_i \to -\infty  ~~~ \text{and} & u_7 & = u_8\: .
\end{aligned}
\end{align}
We used the asymptotic behavior of the hyperbolic gamma function
\begin{align}
	\lim_{z\to\infty}e^{\frac{\pi i}{2}B_{2,2}(z;\omega_{1},\omega_{2})}\gamma^{(2)}(z;\omega_{1},\omega_{2})=1\: \: \text{for} \:  \: \arg{\omega_{2}+\pi}>\arg{z}>\arg{\omega_{1}}   \\
	\lim_{z\to\infty}e^{-\frac{\pi i}{2}B_{2,2}(z;\omega_{1},\omega_{2})}\gamma^{(2)}(z;\omega_{1},\omega_{2})=1 \: \: \text{for} \: \: \arg{\omega_{2}}>\arg{z}>\arg{\omega_{1}-\pi} \; ,\label{asymptotics}
\end{align}
where $\text{Im}(\frac{\omega_{1}}{\omega_{2}})>0$.

One can observe that the limit sends spins $\sigma_2$ and $\sigma_4$ to infinity, and the remaining Boltzmann weights stay alive for $\sigma_1$ and $\sigma_3$ spins with the exchanged spectral parameters in (\ref{factorized1}) and (\ref{factorized2}). We also note that the gauge factors in the star-star relation (\ref{starstar}) disappear in the limit. A similar reduction from the star-star relation to the flipping relation can be investigated for the lens version of the Faddeev-Volkov model \cite{Bozkurt:2020gyy, Catak:2021coz}.

\section{More solutions to the flipping relation}

\subsection{Hyperbolic solution}

We derive an integral transformation formula by the use of a hyperbolic beta integral of Askey-Wilson type \cite{STOKMAN2005119, Ruijsenaars2003, Kels:2018xge, Sarkissian:2020ipg}. The integral identity is introduced as a solution to the decoration transformation. From the gauge theory side, the identity is the result of the equality of the partition functions of the three-dimensional $\mathcal{N} = 2$ supersymmetric dual gauge theories on $S_b^3$ \cite{Hosomichi:2014hja, Willett:2016adv}. 

Two theories have the same flavor symmetry but only one theory possess a $SU(2)$ gauge symmetry. In the theory with gauge degrees of freedom, chiral multiplets transform under the fundamental representation of the gauge group and the flavor group, and the vector multiplet transforms as the adjoint representation of the gauge group. However the dual theory consists of only  seven chiral multiples in the totally antisymmetric tensor representation of the flavor group. This theory is also deduced from similar duality but the global symmetry is $SU(6)$ flavor symmetry. The duality with $SU(6)$ flavor symmetry result in the integral identity which is studied as a solution to the star-triangle relation. Here we show that the Boltzmann weights of this integrable model has a a flipping property.

Then, we deduce the following integral identity by the use of double integral method \cite{Catak:2021coz}
\begin{align}
        \int_{-i\infty}^{i\infty}&\frac{\prod_{j=1}^4\gamma^{(2)}(a_j\pm z;\omega_1,\omega_2)}{\gamma^{(2)}(\pm2z;\omega_1,\omega_2)}\frac{dz}{2i\sqrt{\omega_1\omega_2}}
\nonumber\\ 
&=\frac{\gamma^{(2)}(a_1+a_2;\omega_1,\omega_2)\gamma^{(2)}(a_3+a_4;\omega_1,\omega_2)}{\gamma^{(2)}(\Tilde{a}_1+\Tilde{a}_2;\omega_1,\omega_2)\gamma^{(2)}(\Tilde{a}_3+\Tilde{a}_4;\omega_1,\omega_2)}
\nonumber\\ &\quad \times
      \int _{-\infty}^{\infty} \frac{\prod_{i=1}^4\gamma^{(2)}(\Tilde{a}_i\pm x;\omega_1,\omega_2)}{\gamma^{(2)}(\pm 2x;\omega_1,\omega_2)}\frac{dx}{2i\sqrt{\omega_1\omega_2}}
      \label{flippingequation2}
\end{align}
where $s$ is a chemical potential and
\begin{align}
    \Tilde{a}_{1,2}=a_{1,2}+s \:,\quad
\Tilde{a}_{3,4}=a_{3,4}-s \:.
\end{align}
Under the proper change of the variables,
\begin{align}
    a_{1,2}=-\alpha_1\pm x_1 \:,\quad
a_{3,4}=-\alpha_2\pm x_2\:,\quad
    s=\alpha_1-\alpha_2 \:,
\end{align}
the integral identity (\ref{flippingequation2}) becomes the flipping relation for the Boltzmann weights \cite{Spiridonov:2010em}  
 \begin{align}
    \begin{aligned}
W_\alpha(x_i,x_j)=\gamma^{(2)}(-\alpha \pm x_i\pm x_j ;\omega_1,\omega_2)\:,
\end{aligned}\label{integrableB1}
\end{align}
The integrable spin model has a self-interaction term for each spin
\begin{align}
   S(x_0)= \frac{1}{\gamma^{(2)}(\pm2x_0;\omega_1,\omega_2)}\:.\label{self1}
\end{align}

\subsection{Trigonometric solution}
The trigonometric solutions to the integrability conditions are obtained from the basic hypergeometric integral identities \cite{ Gahramanov:2015cva, Gahramanov:2023lwk}. These integral identities are the partition functions of the $\mathcal{N}=2$ supersymmetric gauge theories on $S^2 \times S^1$. In \cite{Kels:2015bda}, the integrable trigonometric model is investigated as a dimensionally reduced version of the lens elliptic model \cite{Yamazaki:2013nra}, see also \cite{Kels:2017toi, Kels:2017vbc}.

From the supersymmetric gauge theory perspective, the integral identity as a decoration transformation is an equality for partition functions of dual gauge theories. Both dual theories have the $SU(4)$ flavor symmetry, and only one possesses $SU(2)$ gauge symmetry \cite{Gahramanov:2016wxi}. We apply the double integral method and obtain the following
\begin{align} \nonumber
 \sum_{m\in\mathbb{Z}}\oint \frac{dz}{4 \pi i z}\frac{(1-q^{m} z^2)(1-q^{m} z^{-2})}{q^m     z^{4m} }
\prod_{j=1}^4 
\frac{(q^{1+\frac{n_j+m}{2} }/{a_jz},q^{1+\frac{n_j-m}{2} }{z}/{a_j};q)_\infty}
{(q^{\frac{n_j+m}{2} }a_jz,q^{\frac{n_j-m}{2} }{a_j}/{z};q)_\infty}
\nonumber \\ = 
\frac{
\frac{(q^{1+\frac{n_1+n_2}{2}}/a_1a_2;q)_\infty (q^{1+ \frac{ n_3+n_4}{2}}/ a_3 a_4)_{\infty}}{(q^{\frac{n_1+n_2}{2}}a_1a_2;q)_\infty(q^{\frac{ n_3+n_4}{2}} a_3 a_4)_{\infty}}
}{
\frac{(q^{1+ \frac{ \tilde{n}_1+\tilde{n}_2}{2}}/\tilde{a}_1 \tilde{a}_2 )_{\infty}(q^{1+\frac{\tilde{n}_3+\tilde{n}_4}{2}}/\tilde{a}_3\tilde{a}_4;q)_\infty}{( q^{\frac{ \tilde{n}_1+\tilde{n}_2}{2}}\tilde{a}_1 \tilde{a}_2 )_{\infty} (q^{\frac{\tilde{n}_3+\tilde{n}_4}{2}}\tilde{a}_3\tilde{a}_4;q)_\infty}}
\nonumber \\
\sum_{y\in\mathbb{Z}}\oint \frac{dx}{4 \pi i x}\frac{(1-q^{y} x^2)(1-q^{y} x^{-2})}{q^y     x^{4y} }
\prod_{j=1}^4 
\frac{(q^{1+\frac{\tilde{n}_j+y}{2} }/{\tilde{a}_jx},q^{1+\frac{\tilde{n}_j-y}{2} }{x}/{\tilde{a}_j};q)_\infty}
{(q^{\frac{\tilde{n}_j+y}{2} }\tilde{a}_jx,q^{\frac{\tilde{n}_j-y}{2} }{a_j}/{x};q)_\infty} \:,
\label{trigoflipping}
\end{align}
where some fugacities are defined as
\begin{align}
  &\tilde{a}_j=a_js  \:, \quad 
   \tilde{n}_j=n_j+p\:, \quad j=1,2\:, \nonumber \\
   &\tilde{a}_j=a_js^{-1}  \:, \quad 
 \tilde{n}_j=n_j-p \:, \quad j=3,4\:.
\end{align}
If one redefines fugacities and chemical potentials as
\begin{align}
 a_j&=x_1^{\pm1}\alpha_1^{-1}\:, \quad \quad  u_j=\pm m_1-\beta_1\quad 
  j=1,2\:, \nonumber \\
   a_j&=x_2^{\pm1}\alpha_2^{-1}\:, \quad \quad u_j=\pm m_2-\beta_2\quad 
  j=3,4\:,
  \nonumber \\
  s&=\alpha_1\alpha_2^{-1}\:, \quad \quad p=\beta_1-\beta_2\:,
\end{align}
the integral identity (\ref{trigoflipping}) turns into the flipping relation with the Boltzmann weight of the integrable trigonometric model 
\begin{align}
 \begin{aligned}
    W_{\alpha, \beta}(\sigma_i,\sigma_0)=
    \frac{(q^{1+(-\beta_i\pm v_i\pm v_0)/2}(\alpha_i^{-1}x_i^{\pm1}x_0^{\pm1})^{-1};q)_\infty}{(q^{(-\beta_i\pm v_i\pm v_0)/2}\alpha_i^{-1} x_i^{\pm1}x_0^{\pm1};q)_\infty}\;.
    \end{aligned}\label{integrableB3}
\end{align}
And the self-interaction term appears
\begin{align}
    S(\sigma_0)=\frac{(1-q^{m_0} x_0^2)(1-q^{m_0} x_0^{-2})}{q^{m_0}     x_0^{4m_0} }
    \;.\label{self3}
\end{align}

\subsection{Rational solution}
The ordinary limit of the lens hyperbolic gamma function \cite{Eren:2019ibl, Mullahasanoglu:2024stv} is
\begin{equation}
	\lim_{r\to\infty} \gamma_h(z,y;\omega_1,\omega_2)=\Big(\frac{r}{4\pi}\Big)^{\frac{2-2iz}{r}}\frac{\Gamma\left(\frac{iz+y}{2}\right)}{\Gamma\left(1-\frac{iz-y}{2}\right)}\equiv \Big(\frac{r}{4\pi}\Big)^{\frac{2-2iz}{r}}{\bf \Gamma}(z,y)\:.
\end{equation} 
where the complex gamma function ${\bf \Gamma}(z,y)$ is defined as a certain combination of Euler's gamma functions.

This asymptotic property of the lens hyperbolic gamma function helps us to reduce the integral identity (\ref{flippingequation1}) into
\begin{align}
     \sum_{-\infty}^{\infty} \int _{-\infty}^{\infty} dz &
    (z^2+y^2)
    \prod_{i=1}^4\boldsymbol{\Gamma}\left(a_i\pm z ,u_i \pm y \right)\\
    \nonumber & = \frac{\boldsymbol{\Gamma}\left( a_1+a_2,u_1+u_2\right)\boldsymbol{\Gamma}\left( a_3+a_4,u_3+u_4\right)}{\boldsymbol{\Gamma}\left(\Tilde{a}_1+\Tilde{a}_2,\Tilde{u}_1+\Tilde{u}_2\right)\boldsymbol{\Gamma}\left(\Tilde{a}_3+\Tilde{a}_4,\Tilde{u}_3+\Tilde{u}_4\right)}
    \nonumber \\ &\times 
    \sum_{-\infty}^{\infty}\int _{-\infty}^{\infty} dx 
    (x^2+m^2)
    \prod_{i=1}^4 \boldsymbol{\Gamma}\left(\Tilde{a}_i\pm x ,
    \Tilde{u}_i \pm m\right)
    \label{hyper44}
\end{align}
where tilde parameters are
\begin{align}
    \Tilde{a}_{1,2}=a_{1,2}+s \:,\quad
\Tilde{a}_{3,4}=a_{3,4}-s \:,\quad
  \Tilde{u}_{1,2}= u_{1,2}+p \:,\quad
\Tilde{u}_{3,4}=u_{3,4}-p \:.
\end{align}

We want to write the integral identity (\ref{hyper44}) as the flipping relation under the change of variables (\ref{tilde1}) and (\ref{sp}). Then we deduce that the Boltzmann weight introduced in \cite{Kels:2013ola, Kels:2015bda, Eren:2019ibl} solves the flipping relation
 \begin{align}
    \begin{aligned}
        W_{\alpha, \beta}(\sigma_i,\sigma_0)={\bf \Gamma}(-\alpha \pm x_i\pm x_0,-\beta \pm v_i\pm v_0 )
       \:. \label{integrableB4}
    \end{aligned}
\end{align}
The self-interaction term is
\begin{align}
    S(\sigma_0)=x_0^2+v_0^2 \:.\label{self4}
\end{align}

\section{Conclusion}
In this work, we re-derived the lens hyperbolic solution to the flipping relation with a different approach. In \cite{Catak:2024ygo}, the solution is obtained by applying the double integral method to the integral identity, solving the decoration transformation. We obtained the same result by reducing the four fugacities, two spins, from the integral identity for the star-star relation (\ref{starstar}). That is, we explicitly show that the flipping relation is a reduced version of the star-star relation. 

We further establish that integrable lattice spin models constructed via the gauge/YBE correspondence inherently possess a flipping relation. We solve the flipping relation using Boltzmann weights defined by the hyperbolic gamma function, the basic hypergeometric function, and the complex Euler gamma function.

There are classical limits of the star-triangle relation and the star-star relations  \cite{Kels:2020zjn, Kels:2024afm}. As further studies, it would be interesting to see what role the flipping relation plays in the context of consistency equations on a face-centered cubic.

\section*{Acknowledgements}
We thank to Ilmar Gahramanov for his comments on manuscript.
We acknowledge support from the Istanbul Integrability and Stringy Topics Initiative (\href{https://istringy.org/}{istringy.org}). Erdal Catak and Mustafa Mullahasanoglu are supported by the Scientific and Technological Research Council of Turkey (TÜBİTAK) under grant number 122F451.


\bibliographystyle{utphys}
\bibliography{refYBE}

\end{document}